\documentclass[12pt]{article} 

\begin{document}

\begin{center} 
{\Large {\bf The Masses of all the fermions in Supersymmetric 
$SU(3)_{C}\times SU(3)_{L}\times U(1)_{N}$ Model with $R$-Parity Conservation.}}
\end{center}

\begin{center}
M. C. Rodriguez  \\
{\it Grupo de F{\'{\i}}sica Te\'{o}rica e Matem\'{a}tica F\'{\i}sica \\
Departamento de F\'{\i}sica  \\
Universidade Federal Rural do Rio de Janeiro (UFRRJ) \\
BR 465 Km 7, 23890-000 Serop\'{e}dica - RJ \\
Brasil}
\end{center}

\date{\today}

\begin{abstract}
We will get the mass spectrum of all the fermions in the context 
of the supersymmetric $SU(3)_{C}\times SU(3)_{L}\times U(1)_{N}$ 
Model, when all the usual neutral scalars fields 
obtain vacuum expectation values. In this model, as happen in MSSM, we can define a 
$R$-parity conservation, where our superpotential respect the invariance 
of the quantum number, ${\cal F}\equiv B+L$, where $B$ is the baryon 
number while $L$ is the total lepton number. We will present numerical 
predictions for all the fermions of this model and we will show that 
all masses and mixing angles are in agreement with current 
experimental data.
\end{abstract}

PACS number(s): 12.60. Cn, 12.60. Jv



\section{Introduction}

The  current values for the masses of charged fermions are \cite{pdg}
\begin{eqnarray}
m_{e}&=&0.005\;{\mbox GeV}, \,\ m_{\mu}=0.105\;{\mbox GeV}, \,\ 
m_{\tau}=1.77\;{\mbox GeV}, \nonumber \\
m_{u}&\sim&1-5\;{\mbox MeV},\,\ m_{d}\sim3-9\;{\mbox MeV},\;m_{s}\sim75-170\;{\mbox MeV}, \nonumber \\
m_{c}&\sim&1.15-1.35\;{\mbox GeV},\,\ m_{b}\sim4.0-4.4\;{\mbox GeV}.
\label{chargedfermionmasses}
\end{eqnarray}
The value of the top quark mass provided by the Large Hadron Collider (LHC)  is \cite{Sirunyan:2018gqx}  
\begin{equation} 
m_{t}=172.25\pm 0.08\mathrm{(stat.)} \pm 0.62\mathrm{(syst.)}~{\mbox GeV}.
\label{mtexp}
\end{equation}

The charged weak interactions between the quarks and the charged gauge 
boson $W^{\pm}$ is given by \cite{Bilenky:1982ms,Bilenky:1982tw,donoghue,quigg,chengli,peskin}
\begin{eqnarray}
{\cal L}^{qq^{\prime}}_{W}&=&- \frac{g}{\sqrt{2}} 
\left( \overline{u^{\prime}_{iL}}\gamma^{m}d^{\prime}_{jL} W_{m}^{+} + hc \right)=- \frac{g}{\sqrt{2}} 
\left( \overline{u_{iL}} \gamma^{m}d^{\prime \prime}_{jL}W^{+}_{m}+hc \right) , \nonumber \\
d^{\prime \prime}_{jL}&=&V^{CKM}_{jl}d^{\prime}_{lL}, \,\ 
V^{CKM}= \left( S^{u}_{L}\right)^{\dagger}S^{d}_{L}, \,\ i,j=1,2,3,
\label{ckmdef}
\end{eqnarray}
where $V^{CKM}$ is the Cabibbo-Kobayashi-Maskawa (CKM) matrix and it 
is parameterized as
\begin{equation}
V_{CKM}=\left(
\begin{array}[c]{ccc}
V_{ud} & V_{us} & V_{ub}\\
V_{cd} & V_{cs} & V_{cb}\\
V_{td} & V_{ts} & V_{tb}
\end{array}
\right),
\end{equation}
where the matrix element $V_{ij}$ indicates the contribution of quark ($j$) to quark ($i$). 
The experimental values are \cite{pdg}
\begin{equation}
\left(
\begin{array}[c]{ccc}
0.9739-0.9751 & 0.221-0.227 & 0.0029-0.0045\\
0.221-0.227 & 0.9730-0.9744 & 0.039-0.044\\
0.0048-0.014 & 0.037-0.043 & 0.9990-0.9992
\end{array}
\right)  
\label{datackm}
\end{equation}

Today we know neutrinos are massive particles and they can oscillate  \cite{Cribier:2019ckv,bilenkyprehist}. To explain the 
experimental data, we need of at least two massive neutrinos. The theory beyond oscillations can be summarize as: 
the neutrino state was created in the decay
\begin{equation}
  W^{-}\rightarrow l_{i} + \bar{\nu}_{i}, \,\ W^{+}\rightarrow l_{i}^{+} + \nu_{i},
\end{equation}
and it is described by the following lagrangian
\begin{eqnarray}
{\cal L}^{ll}_{W}&=& -
\frac{g}{\sqrt{2}} \sum_{i=1}^{3} \left( \overline{l^{\prime}_{Li}} \gamma^{m}  
\nu^{\prime}_{Li} W_{m}^{-} + hc \right)=-
\frac{g}{\sqrt{2}} \sum_{i,j=1,2,3} 
\left( \overline{l_{Li}} 
\gamma^{m} U_{ij} \nu_{Lj} W_{m}^{-} + hc \right). \nonumber \\
\label{eq2.2}
\end{eqnarray}
We define the lepton mixing matrix as\footnote{In similar way arise the CKM matrix, given by our Eq.(\ref{ckmdef})} 
\begin{equation}
U_{PMNS}= \left( V^{l}_{L} \right)^{\dagger}V^{\nu}_{L}.
\label{pmnsdef}
\end{equation}
The unitary matrix, $U_{PMNS}$, is known as the Pontecorvo-Maki-Nakagawa-Sakata (PMNS)\footnote{About Pontecorvo see 
\cite{bilenkyprehist,Dore:2009bq,Bilenky:2013wna}}. 

The best-fit values at 
$1 \sigma$ error level are summarised as follows \cite{Esteban:2024eli} 
\begin{itemize}
\item Normal Hierarchy 
\begin{eqnarray}
\sin^{2}\theta_{12}&=&\sin^{2} \theta_{solar} = 0.308^{+0.012}_{-0.011}, \nonumber \\
\Delta m^{2}_{21}&=& \Delta m_{solar}^{2} = 7.49^{+0.19}_{-0.19} \times 10^{-5}\,\ {\mbox eV}^{2}\;, \nonumber \\
\sin^{2} \theta_{23} &=& \sin^{2} \theta_{atm} = 0.470^{+0.017}_{-0.013}, \nonumber \\
|\Delta m^{2}_{23}|&=& \Delta |m_{atm}^{2}| = 2.513^{+0.021}_{-0.019} \times 10^{-3}\,\ {\mbox eV}^{2}\;, \nonumber \\
\sin^{2} \theta_{13} &=&\sin^{2} \theta_{CHOOZ} =  0.02215^{+0.00056}_{-0.00058}\;. \nonumber \\
\label{eq1:best-fit_mass}
\end{eqnarray}
\item Inverted Hierarchy 
\end{itemize}
\begin{eqnarray}
\sin^{2}\theta_{12}&=& 0.308^{+0.012}_{-0.011} \,,
\Delta m^{2}_{21}=  7.49^{+0.19}_{-0.19} \times 10^{-5}\,\ {\mbox eV}^{2}\;, \nonumber \\
\sin^{2} \theta_{23} &=& 0.550^{+0.012}_{-0.015}\,,
|\Delta m^{2}_{23}|= 2.484^{+0.020}_{-0.020} \times 10^{-3}\,\ {\mbox eV}^{2}\;, \nonumber \\
\sin^{2} \theta_{13} &=& 0.02231^{+0.00056}_{-0.00058}\;. \nonumber \\
\label{eq1:best-fit_massin}
\end{eqnarray}

Supersymmetry, it is more known as SUSY, arose in theoretical papers more 
than 30 years ago independently by 
Golfand and Likhtman \cite{gl}\footnote{They constructed the first four-dimensional field 
theory with Supersymmetry, (massive) Quantum Electrodynamics (SQED) of spinors and scalars. 
We will call it as {\bf Golfand-Likhtman Model} as suggested by Shiffman \cite{Shifman:2015tka}.}. The Volkov-Akulov article, 
however, deals only with fermions and its name is ``{\bf Is the Neutrino a Goldstone Particle?}"
\footnote{It is the base of Supergravity Theory.} \cite{va} and 
Wess and Zumino \cite{wz}. For interested people interested in the history of supersymmetry see 
\cite{volkov1,Likhtman:2001st}. The history of the first Supersymmetric versions of the Standard Model can be seen in \cite{fay1,marcoshist}. There are good reviews of supersymmetric theories \cite{ogievetski,wb,MullerKirsten:1986cw}.

Since that time there appeared thousands of papers and the reason for this remarkable is due the fact that 
there are a number of theoretical and phenomenological issues that the Standard Model (SM) fails to address appropriately 
\cite{Chung:2003fi,dress,Baer:2006rs,Aitchison:2005cf,mssm,Simonsen:1995cf,Kuroda:1999ks,kraml} 
\begin{itemize}
\item Unification with Gravity: This point arises from the fact that SUSY Algebra being a generalization of
Poincar\'e Algebra \cite{wb,dress,Baer:2006rs,Aitchison:2005cf}
\begin{equation}
\{Q_\alpha, \bar{Q}_{\dot{\alpha}}\}=2\sigma_{\alpha \dot{\alpha}}^{m}P_{m},
\end{equation}
when we make SUSY local, one obtain Supergravity Theory~\cite{sugra}.
\item Unification of Gauge Couplings: According to hypothesis of Grand Unification Theory (GUT) all gauge couplings change with energy. In the Minimal Supersymmetric Standard Model (MSSM) the slopes of Renormalization Group Equation (RGE) curves about the 
running of behaviour of gauge couplings can achieve perfect unification \cite{ABF,Ellis:1990wk,Langacker:1991an}.
\item Hierarchy Problem: The Supersymmetric Theories automatically cancels all
quadratic corrections in all orders of perturbation theory due to
the contributions of superpartners of the ordinary particles.  A light Higgs sector ($M^{2}_{H}\ll M^{2}_{Pl}$) is natural in a 
softly broken SUSY scenarios where the quadratically divergent loop contribution cancel leaving a finite correction of the form 
\begin{equation}
\delta M^{2}_{H}= {\cal O}\left( \frac{\alpha}{\pi} \right) \left(
M^{2}_{B}- M^{2}_{F}\right) ,
\end{equation}
where $M_{B,F}$ are the masses of the bosonic and fermionic partner
particles circulating in the loops. When we consider 
scenarios with soft SUSY breakings terms we have the following constraints
\begin{equation}
M^{2}_{B}-M^{2}_{F}\simeq {\cal O}(M_{W})= M_{SUSY}^{2},
\end{equation}
which  should not be very large ($\leq$ 1 TeV) to make the fine-tuning natural.  
\item Electroweak Symmetry Breaking (EWSB): The ``running" of the Higgs masses, using the RGE of MSSM, leads to the phenomenon known as radiative Electroweak Symmetry Breaking. Indeed, the mass parameters from the Higgs
potential $m_{1}^{2}$ and $m_{2}^{2}$ (or one of them) decrease while
running from the GUT scale to the scale $M_{Z}$ may even change the
sign.  Thus the breaking of the Electroweak Symmetry is not introduced by brute force as in 
the SM, but appears naturally from the radiative corrections \cite{running}.
\end{itemize}

The main sucess of SUSY  is to solve all the problems listed above. However, SUSY has also 
made several correct predictions \cite{Chung:2003fi}
\begin{itemize}
\item SUSY predicted in the early 1980s that the top quark would be heavy \cite{Ibanez:wd,Pendleton:as}.
\item SUSY GUT theories with a high fundamental scale accurately predicted the present experimental value of $\sin^{2} \theta_{W}$ 
before it was mesured \cite{Dimopoulos:1981yj,Dimopoulos:1981zb,Ibanez:yh,Einhorn:1981sx}.
\item SUSY requires a light Higgs boson to exist \cite{INO82a,Kane:1992kq,Espinosa:1992hp,haber2,Djouadi:2005gj,Casas:1994us}, its 
experimental value is $M_{H} = 125.09\pm0.21\,\mathrm{(stat.)}\pm0.11\,\mathrm{(syst.)}~\mathrm{GeV}$ \cite{Aad:2015zhl}.
\end{itemize}
Together these success provide powerful indirect evidence that low energy SUSY is indeed part 
of correct description of nature. The Minimal Supersymmetric Standard Model (MSSM) 
\cite{dress,Baer:2006rs,Aitchison:2005cf,mssm,Simonsen:1995cf,Kuroda:1999ks}\footnote{About the 
history of MSSM, see e. g. \cite{fay1,marcoshist}}  have, also viable candidates to 
be the Dark Matter on this case it can be the lightest neutralino, 
lighest sneutrino\footnote{Unfortunatelly this particle have been ruled out by the combination of collider experiment as LEP and direct 
searches for cosmological relics as discussed in \cite{hillwalkers}.} and the gravitino, more details about SUSY Dark Matter Candidates 
is presented in nice way in \cite{Ellis:2010kf}.

The MSSM is a good candidate to be the physics beyond the SM. In this 
model we can solve the hierarchy problem as well to explain the Higgs 
Masses
\footnote{The current status of the search for supersymmetry is presented in 
reference \cite{gladkaza,GAMBIT:2018gjo}.} \cite{kazakov1,Rodriguez:2019mwf}. The experimental 
lower limits in neutralinos and charginos are \cite{GAMBIT:2018gjo}
\begin{eqnarray}
155&<&m_{\tilde{\chi}^{0}_{1}}> 8 \,\, {\mbox GeV}, \nonumber \\
259&<&m_{\tilde{\chi}^{\pm}_{1}}> 104 \,\, {\mbox GeV}.
\label{limitesnoMSSM}
\end{eqnarray}

Models with the gauge symmetry \cite{singer,ppf,331rh} 
\begin{equation}
SU(3)_{C} \times SU(3)_{L} \times U(1)_{N}
\end{equation} 
are known as $331$ for short. They are interesting  
possibilities for the physics at the TeV scale \cite{Pleitez:1994pu,Ponce:2001jn}. It is a subgroup of 
unification group $E_{6}$ \cite{Sanchez:2001ua} and it is also 
an $SU(6) \times U(1)_{X}$  \cite{Martinez:2001mu}, 
$SU(8)$ \cite{Martinez:2011iq} and $SU(15)$ \cite{Frampton:1989fu} subgroups.

Although this model coincides at low energies with the SM it explains 
some fundamental questions that are accommodated, but not explained, in the SM. These questions are
\begin{enumerate}
\item The family number must be a multiple of three in order to 
cancel anomalies \cite{ppf,331rh}\footnote{This model is anomaly free if we have equal number 
of triplets and anti-triplets, counting the color of $SU(3)_{C}$, and furthermore requiring the 
sum of all fermion charges to vanish \cite{ppf}.}.
\item Why $\sin^{2} \theta_{W}<\frac{1}{4}$ is observed \cite{ppf}.
\item It is the simplest model that includes bileptons of both types: Scalars and Vectors ones \cite{Cuypers:1996ia};
\item The models have a scalar sector similar to the Two Higgs Doublets Models 
\cite{Liu:1993gy,Liu:1993fwa,Montero:1999mc,DeConto:2015eia}.
\item It solves the strong $CP$-Problem \cite{pal1}.
\item The model has several sources of $CP$-Violation \cite{Montero:1998nf,Montero:1998yw,Montero:2005yb}.
\end{enumerate}

There are several distinct possible models based on this gauge symmetry. The reason for this is that the electric charge 
operator, in the $SU(3)_{L}$ generators, is defined as
\begin{eqnarray}
\frac{Q}{e}&=& \frac{1}{2}(\lambda_{3}- \vartheta \lambda_{8})+N \,\ I_{3 \times 3}, 
\label{co}
\end{eqnarray}
where the $\vartheta$ and $N$ are parameters defining differents representation contents and $\lambda_{3}$, $\lambda_{8}$ are the diagonal generators of $SU(3)$ given by
\begin{eqnarray}
\lambda_{3}= \left( \begin{array}{ccc} 
+1 & 0 & 0\\ 
0 & -1 & 0 \\
0 & 0 & 0          
\end{array} \right), \,\
\lambda_{8}= \frac{1}{\sqrt{3}} \left( \begin{array}{ccc} 
+1 & 0 & 0\\ 
0 & +1 & 0 \\
0 & 0 & -2          
\end{array} \right).
\end{eqnarray}

One of the possible and well-studied models in the literature is the 
model proposed by Pleitez-Pisano~\cite{ppf}, where we chose 
$\vartheta=\sqrt3$. In this case, Eq.(\ref{co}) become
\begin{eqnarray}
\frac{Q}{e}&=& \left( \begin{array}{ccc} 
N & 0 & 0 \\ 
0 & N-1 & 0 \\
0 & 0 & N+1          
\end{array} \right).
\label{mppf}
\end{eqnarray} 
The lepton sector is exactly the same as in the Standard Model (SM) \cite{sg,Kronfeld:2010bx} but now there is a symmetry, at 
large energies among, say the electron ($e^{-}$), its neutrino ($\nu_{e}$) and the positron ($e^{+}$) and in this model 
the leptons are \cite{ppf}
\begin{eqnarray}
L^{PP}_{iL} &=&\left( \begin{array}{c} 
\nu_{i} \\ 
l_{i} \\
l^{c}_{i}          
\end{array} \right)_{L} \sim ({\bf1},{\bf3},0), \,\ i= 1,2,3.
\label{tripPP}
\end{eqnarray}
this model is known as M331 and its supersymmetric version has already been considered in 
Refs.~\cite{ema1,pal2,331susy1,mcr,Rodriguez:2005jt}, and we will call it as 
MSUSY331. There are another interesting possibility propused by Pleitez-Tonasse \cite{Pleitez:1992xh}
\begin{eqnarray}
L^{PT}_{iL} &=&\left( \begin{array}{c} 
\nu_{i} \\ 
l_{i} \\
E^{+}_{i}          
\end{array} \right)_{L} \sim ({\bf1},{\bf3},0), \,\ i= 1,2,3.
\label{tripPT}
\end{eqnarray}
where $E^{+}$ 
is an extra charged leptons which do not mix with the known 
leptons \cite{Pleitez:1992xh,Pleitez:1993gc}. We want to remind that there is no right-handed 
(RH) neutrino in both  model presented above. The $331$ model of 
Refs.~\cite{331rh}, is know as $331$ model with right-handed neutrinos. we 
define $\vartheta=(1/\sqrt{3})$ and we get
\begin{eqnarray}
\frac{Q}{e}&=& \left( \begin{array}{ccc} 
N+ \frac{1}{3} & 0 & 0 \\ 
0 & N- \frac{2}{3} & 0 \\
0 & 0 & N+ \frac{1}{3}          
\end{array} \right),
\label{331rhn}
\end{eqnarray} 
and in the lepton sector, we have again the electron ($e^{-}$), its neutrino ($\nu_{e}$) and 
the anti-neutrino ($\nu^{c}_{e}$) \cite{331rh}
\begin{eqnarray}
L^{331rh}_{iL} &=&\left( \begin{array}{c} 
\nu_{i} \\ 
l_{i} \\
\nu^{c}_{i}          
\end{array} \right)_{L} \sim ({\bf1},{\bf3},0), \,\ i= 1,2,3.
\label{trip331rh}
\end{eqnarray}
the supersymmetric version of this model was 
built in \cite{331susy2,huong,Huong:2012pg}. Clearly, we can also have models similar to 
the two districts above, having heavy leptons, such as $E^{+}$ instead of $e^{+}$ or even $N^{c}$ replacing $\nu^{c}$, for more details see \cite{Pleitez:1994pu}. There are also anoher interesting possibilities as 
the simplest 331 model \cite{Dong:2014esa}, the flipped 331 model \cite{Fonseca:2016tbn,Hong:2020qxc}.   There are anothers versions, please see \cite{Pleitez:1994pu,Long:2024gyy,Diaz:2004fs,CarcamoHernandez:2005ka}.
The $331$ models are interesting  possibilities for the physics at the TeV scale, 
some phenomenological analyses was presented by references 
\cite{Gonzalez-Sprinberg:2005ayv,Das:1998qh,Das:1999hn,Buras:2012dp,Buras:2014yna,Cao:2016uur,Buras:2021rdg}. 

The quark sector, we introduce two generations, in the 
anti-triplet representation of $SU(3)_{L}$\footnote{In parenthesis it appears the transformations properties under 
the respective factors $(SU(3)_{C},SU(3)_{L},U(1)_{N})$.} \cite{ppf}
\begin{equation}
\begin{array}{cc}  
Q_{1L} =\left( \begin{array}{c} 
d_{1} \\ 
u_{1} \\
j_{1}          \end{array} \right)_{L},\quad  
 
Q_{2L} =\left( \begin{array}{c} 
d_{2} \\ 
u_{2} \\
j_{2}          \end{array} \right)_{L} 
\sim \left({\bf3},{\bf\bar{3}},-\frac{1}{3}\right) , 
\end{array}
\label{q23l}
\end{equation} 
and its singlets fields which we can write as follows
\begin{equation}
u_{1R} \,\ , u_{2R} \sim \left({\bf3},{\bf1},\frac{2}{3}\right),\quad
d_{1R} \,\ , d_{2R} \sim \left({\bf3},{\bf1},- \frac{1}{3}\right), \quad
j_{1R} \,\ , j_{2R} \,\ \sim \left({\bf3},{\bf1},- \frac{4}{3} \right) 
\,\ . 
\label{q23r}
\end{equation} 

We place one familly in the triplet representation of 
$SU(3)_{L}$ \cite{ppf}
\begin{eqnarray}
Q_{3L} &=&\left( \begin{array}{c} 
u_{3} \\ 
d_{3} \\
J          
\end{array} \right)_{L} \sim \left({\bf3},{\bf3},\frac{2}{3}\right) \,\ , 
\label{q1l}
\end{eqnarray}
and the respective singlets are given by
\begin{eqnarray}
u_{3R} &\sim& \left({\bf3},{\bf1},\frac{2}{3}\right),\quad
d_{3R} \sim \left({\bf3},{\bf1},-\frac{1}{3}\right),\quad  
J_{R} \sim \left({\bf3},{\bf1},\frac{5}{3}\right). \nonumber \\
\label{q1r}
\end{eqnarray}
The new quarks\footnote{New compared with the SM.} $j_{\alpha}$ and $J$ are the third component 
of the representation of triplet representation (${\bf 3}$) and anti-triplet representation 
$({\bf \bar{3}})$ of $SU(3)_{L}$ and their charges are $(-4/3)$ and $(5/3)$, respectivelly.

The fact that one family has a different transformation from the other two, could be a possible explanation for why the third family of quarks is much more massive compared to the others two \cite{Pleitez:1994pu}, we will return to this point below. We would like to 
note that this choice for the quarks, defined in Eqs.(\ref{q23l},\ref{q1l}), is not unique; we could to replace the triplets for 
anti-triplets \cite{Diaz:2004fs,CarcamoHernandez:2005ka,Buras:2014yna}. 

In order to generate masses for those quarks, defined in Eqs.(\ref{q23l},\ref{q1l}), we have to introduce the 
following scalars fields
\begin{eqnarray} 
\eta &=&\left( \begin{array}{c} 
\eta^{0} \\ 
\eta^{-}_{1} \\
\eta^{+}_{2}          \end{array} \right) \sim 
({\bf1},{\bf3},0),\quad 
\rho =\left( \begin{array}{c} 
\rho^{+} \\ 
\rho^{0} \\
\rho^{++}          \end{array} \right) \sim 
({\bf1},{\bf3},+1), \nonumber \\
\chi &=& \left( \begin{array}{c} 
\chi^{-} \\ 
\chi^{--} \\
\chi^{0}          \end{array} \right) \sim 
({\bf1},{\bf3},-1),
\label{3t} 
\end{eqnarray}
using those scalars the Yukawa mass term for the quarks are given by \cite{ppf} 
\begin{eqnarray}
{\cal L}^{{\mbox Y}}&=&
\sum_{\alpha =1}^{2}\bar{Q}_{\alpha L}\sum_{i=1}^{3} \left(
F_{\alpha i}u_{i R}\rho^{*} + 
\tilde{F}_{\alpha i}d_{i R}\eta^{*} \right)+ 
\sum_{\alpha =1}^{2}\sum_{\beta =1}^{2}\lambda^{\prime}_{\alpha \beta} \bar{Q}_{\alpha L}j_{\beta R}\chi^{*} \nonumber \\
&+& \bar{Q}_{3L}\sum_{i=1}^{3} \left(
G_{3 i}u_{i R}\eta + 
\tilde{G}_{3 i}d_{i R}\rho \right) + 
\lambda_{J} \bar{Q}_{3L}J_{R}\chi +hc. \nonumber \\
\label{yukawaquarks}
\end{eqnarray}
The masses of quarks in this model is presented \cite{ppf}. The masses of the exotic 
quarks are proportional to $v_{\chi}$, therefore their masses 
are ${\cal O}(1 TeV)$ and they can be discover in Large Hadron Collider (LHC) 
\cite{Gonzalez-Sprinberg:2005ayv,
Das:1998qh,Das:1999hn,Garberson:2013jz,dutta,Rodriguez:2010tn}.

The new exotic quarks, $J$, $j_{1}$ and $j_{2}$, may be discovered by LHC the  thought 
$pp$ collisions, via the following subprocess\footnote{I want to thank Alexander S. Belyaev, who brought this process to my 
attention for these processes, but unfortunately we were unable to publish this study together.}
\begin{eqnarray}
g+d \to U^{--}+J, \,\
g+u \to  U^{--}+j_{\alpha}, 
\label{intersting}
\end{eqnarray} 
its signature is $llXX$ and it can be descted at LHC if they really exist in 
nature \cite{dutta,Rodriguez:2010tn}. There is the following limits in the mass 
of this particle \cite{Gonzalez-Sprinberg:2005ayv}
\begin{equation}
1500 \leq M_{J} \leq 4000, {\mbox GeV},
\label{constraintinMJ}
\end{equation}
it is in agreement with the results presented by the followings authors 
\cite{Das:1999hn,Cao:2016uur,Garberson:2013jz}
\begin{equation}
M_{J}>600, {\mbox GeV}.
\label{vinculochines}
\end{equation}

The Yukawa term for leptons is written as \cite{ppf}
\begin{eqnarray}
{\cal L}^{{\mbox Y}}_{\eta}&=&- \frac{1}{2}\sum_{i=1}^{3}\sum_{j=1}^{3} G^{\eta}_{ij}\epsilon_{abd}
\overline{(L^{a}_{iL})^{c}}L^{b}_{jL}\eta^{d}+hc,
\label{yukawaeta}
\end{eqnarray}
the Yukawa parameter $G^{\eta}_{ij}$, is a matrix $3 \times 3$, and anti-symmetric 
when exchanging family indices $i,j$ and its eigenvalues have 
the following form \cite{ppf}
\begin{equation}
\left( \begin{array}{ccc} 
0 & 0 & 0 \\
0 & M & 0 \\
0 & 0 &- M        
\end{array} \right),
\end{equation}
this means that one charged lepton is massless and the other two are degenerate 
in masses, at least at tree level. 

The simplest solution to correct this problem is to introduce the anti-sextet \cite{ppf}
\begin{equation}
S = \left( \begin{array}{ccc} 
\sigma^{0}_{1}& 
\frac{h^{+}_{2}}{ \sqrt{2}}& \frac{h^{-}_{1}}{ \sqrt{2}} \\ 
\frac{h^{+}_{2}}{ \sqrt{2}}& H^{++}_{1}& \frac{ \sigma^{0}_{2}}{ \sqrt{2}} \\
 \frac{h^{-}_{1}}{ \sqrt{2}}& 
\frac{\sigma^{0}_{2}}{ \sqrt{2}}&  H^{--}_{2}        
\end{array} \right) \sim ({\bf1},{\bf \bar{6}},0),
\label{antisextetdef}
\end{equation}
with this new scalar, we can consider the following Yukawa term
\begin{eqnarray}
{\cal L}^{{\mbox Y}}_{S}&=&- \frac{1}{2}\sum_{i=1}^{3}\sum_{j=1}^{3} G^{S}_{ij}
\overline{(L_{iL})^{c}}L_{jL}S^{ij}+hc,
\label{yukawaS}
\end{eqnarray}
the Yukawa coupling $G^{S}_{ij}$ is anti-symmetric in to change the 
indices $i$ and $j$ \cite{ppf}.

\section{Minimal Supersymmetric $331$ Model.}
\label{msusy1}

We introduce the following chiral superfield for the leptons in the triplet representation of 
$SU(3)_{L}$ and we can decompose those fields in 
\begin{equation}
\left( SU(3)_{C} \times SU(3)_{L}\times U(1)_{N} \right) \supset 
\left( SU(3)_{C} \times SU(2)_{L}\times U(1)_{Y} \right)
\end{equation} 
in the following way \cite{331susy1,mcr,Rodriguez:2010tn,Rodriguez:2022hsj,Rodriguez:2024rvf}
\begin{eqnarray}
\hat{L}_{iL}&=&\left( 
\begin{array}{c}
\hat{l}^{D}_{iL} \\ 
\hat{l}^{c}_{iL}         
\end{array} 
\right) \sim \left( {\bf 1},{\bf 3},0 \right),  \,\ i=1,2,3, 
\nonumber \\
\hat{l}^{D}_{iL}&=&\left( 
\begin{array}{c}
\hat{\nu}_{iL} \\ 
\hat{l}_{iL}         
\end{array} 
\right) \sim \left( {\bf 1},{\bf 2},-1 \right),  \,\
\hat{l}^{c}_{iL}\sim \left( {\bf 1},{\bf 1},2 \right),
\label{quiralLMSUSY331}
\end{eqnarray}
as in the SM, we do not introduce right-handed neutrinos, 
$\left[ \nu_{iR}\sim \left( {\bf 1},{\bf 1},0 \right) \right]$, in this model in 
agreement with the two component theory of neutrino theory of 
Landau \cite{Landau}, Lee and Yang \cite{LeeYang} and Salam \cite{Salam} 
from 1957, later the helicity of neutrinos was measured experimentally in 
1958 and it confirmed this hypothesis \cite{Goldhaber:1958nb}. Those right-handed neutrinos 
are known as sterile neutrinos 
\cite{Volkas:2001zb,Gonzalez-Garcia:2022pbf,Bilenky:2012qb,Petcov:2019yud}. The particle content of 
each chiral superfield and anti-chiral supermultiplet is presented in the 
Tabs.(\ref{lfermionnmssm},\ref{rfermionnmssm}), 
respectively. 

\begin{table}[h]
\begin{center}
\begin{tabular}{|c|c|c|}
\hline 
$\mbox{ Chiral Superfield} $ & $\mbox{ Fermion} $ & $\mbox{ Scalar} $ \\
\hline
$\hat{L}_{iL}=( \hat{l}^{D}_{iL}, \hat{l}^{c}_{iL})^{T}$ & 
$L_{iL}=( l^{D}_{iL},l^{c}_{iL})^{T}$ & 
$\tilde{L}_{iL}=( \tilde{l}^{D}_{iL}, \tilde{l}^{c}_{iL})^{T}$ \\
\hline
$\hat{Q}_{3L}= \left( \hat{q}^{D}_{3L}, \hat{J}_{L} \right)^{T}$ & 
$Q_{3L}=( q^{D}_{3L},J_{L})^{T}$ & 
$\tilde{Q}_{3L}=( \tilde{q}^{D}_{3L}, \tilde{J}_{L})^{T}$ \\ 
\hline
$\hat{\eta} = \left( \hat{\Phi}_{\eta}, \hat{\eta}^{+}_{2} \right)^{T}$ &
$\tilde{\eta} = \left( \tilde{\Phi}_{\eta}, \tilde{\eta}^{+}_{2} \right)^{T}$ &
$\eta = \left( \Phi_{\eta}, \eta^{+}_{2} \right)^{T}$ \\
\hline
$\hat{\rho} = \left( \hat{\Phi}_{\rho}, \hat{\rho}^{++} \right)$ &
$\tilde{\rho} = \left( \tilde{\Phi}_{\rho}, \tilde{\rho}^{++} \right)$ &
$\rho = \left( \Phi_{\rho}, \rho^{++} \right)$ \\
\hline
$\hat{\chi} = \left( \hat{\Phi}_{\chi}, \hat{\chi}^{0} \right)$ &
$\tilde{\chi} = \left( \tilde{\Phi}_{\chi}, \tilde{\chi}^{0} \right)$ &
$\chi = \left( \Phi_{\chi}, \chi^{0} \right)$ \\
\hline
$\hat{S}^{\prime}= \left( 
\begin{array}{cc} 
\hat{T}^{\prime} & \frac{\hat{\Phi}^{\prime}_{S}}{\sqrt{2}} \\ 
\frac{\hat{\Phi}^{\prime T}_{S}}{\sqrt{2}} & \hat{H}^{\prime ++}_{2}          
\end{array} \right)$ &
$\tilde{S}^{\prime}= \left( 
\begin{array}{cc} 
\tilde{T}^{\prime} & \frac{\tilde{\Phi}^{\prime}_{S}}{\sqrt{2}} \\ 
\frac{\tilde{\Phi}^{\prime T}_{S}}{\sqrt{2}} & \tilde{H}^{\prime ++}_{2}          
\end{array} \right)$ & 
$S^{\prime}= \left( 
\begin{array}{cc} 
T^{\prime} & \frac{\Phi^{\prime}_{S}}{\sqrt{2}} \\ 
\frac{\Phi^{\prime T}_{S}}{\sqrt{2}} & H^{\prime ++}_{2}          
\end{array} \right)$ \\
\hline
\end{tabular}
\end{center}
\caption{Particle content in the chiral superfields in MSUSY331 
and we neglected the color indices and $i=1,2,3$.}
\label{lfermionnmssm}
\end{table} 

\begin{table}[h]
\begin{center}
\begin{tabular}{|c|c|c|}
\hline 
$\mbox{ Anti-Chiral Superfield} $ & $\mbox{ Fermion} $ & $\mbox{ Scalar} $  \\
\hline
$\hat{l}^{c}_{iL}$ & $l^{c}_{iL}\equiv \bar{l}_{iR}$ & $\tilde{l}^{c}_{iL}$ \\
\hline
$\hat{Q}_{\alpha L}=( \hat{q}^{D}_{\alpha L}, \hat{j}_{\alpha L})^{T}$ & 
$Q_{\alpha L}=( q^{D}_{\alpha L},j_{\alpha L})^{T}$ & 
$\tilde{Q}_{\alpha L}=
( \tilde{q}^{D}_{\alpha L}, \tilde{j}_{\alpha L})^{T}$ \\
\hline
$\hat{u}^{c}_{iL}$ & $u^{c}_{iL}\equiv \bar{u}_{iR}$ & 
$\tilde{u}^{c}_{iL}$ \\ 
\hline
$\hat{d}^{c}_{iL}$ & 
$d^{c}_{iL}\equiv \bar{d}_{iR}$ & 
$\tilde{d}^{c}_{iL}$ \\ 
\hline
$\hat{j}^{c}_{\alpha L}$ & 
$j^{c}_{\alpha L}\equiv \bar{j}_{\alpha R}$ & 
$\tilde{j}^{c}_{\alpha L}$   \\
\hline
$\hat{J}^{c}_{L}$ & 
$J^{c}_{L}\equiv \bar{J}_{R}$ & 
$\tilde{J}^{c}_{L}$   \\
\hline
$\hat{\eta}^{\prime} = \left( \hat{\Phi}^{\prime}_{\eta}, \hat{\eta}^{\prime -}_{2} \right)^{T}$ &
$\tilde{\eta}^{\prime} = \left( \tilde{\Phi}^{\prime}_{\eta}, \tilde{\eta}^{\prime -}_{2} \right)^{T}$ &
$\eta^{\prime} = \left( \Phi^{\prime}_{\eta}, \eta^{\prime -}_{2} \right)^{T}$ \\
\hline
$\hat{\rho}^{\prime}= \left( \hat{\Phi}^{\prime}_{\rho}, \hat{\rho}^{\prime --} \right)^{T}$ &
$\tilde{\rho}^{\prime}= \left( \tilde{\Phi}^{\prime}_{\rho}, \tilde{\rho}^{\prime --} \right)^{T}$ & 
$\rho^{\prime} = \left( \Phi^{\prime}_{\rho}, \rho^{\prime --} \right)^{T}$ \\
\hline
$\hat{\chi}^{\prime}= \left( \hat{\Phi}^{\prime}_{\chi}, \hat{\chi}^{\prime 0} \right)^{T}$ &
$\tilde{\chi}^{\prime}= \left( \tilde{\Phi}^{\prime}_{\chi}, \tilde{\chi}^{\prime 0} \right)^{T}$ &
$\chi^{\prime} = \left( \Phi^{\prime}_{\chi}, \chi^{\prime 0} \right)^{T}$ \\
\hline
$\hat{S}= \left( 
\begin{array}{cc} 
\hat{T} & \frac{\hat{\Phi}_{S}}{\sqrt{2}} \\ 
\frac{\hat{\Phi}^{T}_{S}}{\sqrt{2}} & \hat{H}^{--}_{2}          
\end{array} \right)$ & 
$\tilde{S}= \left( 
\begin{array}{cc} 
\tilde{T} & \frac{\tilde{\Phi}_{S}}{\sqrt{2}} \\ 
\frac{\tilde{\Phi}^{T}_{S}}{\sqrt{2}} & \tilde{H}^{--}_{2}          
\end{array} \right)$ &
$S= \left( 
\begin{array}{cc} 
T & \frac{\Phi_{S}}{\sqrt{2}} \\ 
\frac{\Phi^{T}_{S}}{\sqrt{2}} & H^{--}_{2}          
\end{array} \right)$ \\
\hline
\end{tabular}
\end{center}
\caption{Particle content in the anti-chiral superfields in MSUSY331 and 
$\alpha , \beta =1,2$, $i=1,2,3$.}
\label{rfermionnmssm}
\end{table} 

The quarks are
\begin{eqnarray}
\hat{Q}_{\alpha L}&=&\left( 
\begin{array}{c}
\hat{q}^{D}_{\alpha L} \\ 
\hat{j}_{\alpha L}         
\end{array} 
\right) \sim \left( {\bf 3},{\bf 3},- \frac{1}{3} \right),  \,\ \alpha =1,2, \nonumber \\  
\hat{q}^{D}_{\alpha L}&=&\left( 
\begin{array}{c}
\hat{d}_{\alpha L} \\ 
\hat{u}_{\alpha L}         
\end{array} 
\right) \sim \left( {\bf 3},{\bf 2}, \frac{1}{3} \right),  \,\
\hat{j}_{\alpha L}\sim \left( {\bf 3},{\bf 1},- \frac{8}{3} \right), \nonumber \\
\hat{Q}_{3L}&=&\left( 
\begin{array}{c}
\hat{q}^{D}_{3L} \\ 
\hat{J}_{L}         
\end{array} 
\right) \sim \left( {\bf 3},{\bf \bar{3}}, \frac{2}{3} \right), 
\nonumber \\
\hat{q}^{D}_{3L}&=&\left( 
\begin{array}{c}
\hat{u}_{3L} \\ 
\hat{d}_{3L}         
\end{array} 
\right) \sim \left( {\bf 3},{\bf 2}, \frac{1}{3} \right),  \,\
\hat{J}_{L}\sim \left( {\bf 3},{\bf 1}, \frac{10}{3} \right).
\label{quiral12QMSUSY331in321}
\end{eqnarray}
The new quarks are singlet under the $SU(2)_{L} \times U(1)_{Y}$, see also the comments below our Eq.(\ref{q1r}).

We have also to introduce the following singlets
\begin{eqnarray}
\hat{u}^{c}_{\alpha L} &\sim& 
\left( {\bf \bar{3}},{\bf 1},- \frac{2}{3} \right),  \,\
\hat{d}^{c}_{\alpha L} \sim 
\left( {\bf \bar{3}},{\bf 1}, \frac{1}{3} \right),  \,\
\hat{j}^{c}_{\alpha L} \sim 
\left( {\bf \bar{3}},{\bf 1}, \frac{4}{3} \right), \nonumber \\
\hat{u}^{c}_{3L} &\sim& 
\left( {\bf \bar{3}},{\bf 1},- \frac{2}{3} \right),  \,\
\hat{d}^{c}_{3L} \sim 
\left( {\bf \bar{3}},{\bf 1}, \frac{1}{3} \right),  \,\
\hat{J}^{c}_{L} \sim 
\left( {\bf \bar{3}},{\bf 1},- \frac{5}{3} \right).
\label{quiral12QMSUSY331rh}
\end{eqnarray}

The scalars of this model are \cite{331susy1,mcr,Rodriguez:2022hsj,Rodriguez:2024rvf}
\begin{eqnarray}
\eta &=& \left( 
\begin{array}{c} 
\Phi_{\eta} \\ 
\eta^{+}_{2}          
\end{array} \right) 
\sim ({\bf1},{\bf3},0), \,\ 
\rho = \left( 
\begin{array}{c} 
\Phi_{\rho} \\ 
\rho^{++}          
\end{array} \right) 
\sim ({\bf 1},{\bf 3},+1),  \nonumber \\
\Phi_{\eta} &=& \left( 
\begin{array}{c} 
\eta^{0} \\ 
\eta^{-}_{1}          
\end{array} \right) 
\sim ({\bf1},{\bf2},-1),\quad 
\eta^{+}_{2} 
\sim ({\bf 1},{\bf 1},+2),  \nonumber \\
\Phi_{\rho} &=& \left( 
\begin{array}{c} 
\rho^{+} \\ 
\rho^{0}          
\end{array} \right) 
\sim ({\bf 1},{\bf 2},+1), \quad
\rho^{++} \sim ({\bf 1},{\bf 1},+4), \nonumber \\
\chi &=& \left( 
\begin{array}{c} 
\Phi_{\chi} \\ 
\chi^{0}          
\end{array} \right) 
\sim ({\bf 1},{\bf 3},-1), \,\ 
\Phi_{\chi} = \left( 
\begin{array}{c} 
\chi^{-} \\ 
\chi^{--}          
\end{array} \right) 
\sim ({\bf 1},{\bf 2},-3),  \nonumber \\
S&=& \left( 
\begin{array}{cc} 
T & \frac{\Phi_{S}}{\sqrt{2}} \\ 
\frac{\Phi^{T}_{S}}{\sqrt{2}} & H^{--}_{2}          
\end{array} \right)
\sim ({\bf 1},{\bf {\bf \bar{6}}},0), \quad
\chi^{0}\sim ({\bf 1},{\bf 1},0), 
\nonumber \\
\Phi_{S}&=& \left( 
\begin{array}{c} 
h^{+}_{2} \\ 
\sigma^{0}_{2}          
\end{array} \right)
\sim ({\bf 1},{\bf 2},+1), \quad 
H^{--}_{2} \sim ({\bf 1},{\bf 1},+4), 
\label{22HM}
\end{eqnarray}
where $\Phi_{S}$ is the Higgs doublet boson of SM and 
$H^{--}_{2}$ appear in the model of Babu presented in reference 
\cite{babuh2mm}. The triplet \cite{Gelmini:1980re}
\begin{eqnarray}
T&=&\left( \begin{array}{cc} 
\sigma^{0}_{1} & \frac{h^{+}_{1}}{\sqrt{2}} \\ 
\frac{h^{+}_{1}}{\sqrt{2}} & H^{--}_{1} \\
\end{array} \right)
\sim ({\bf 1},{\bf 3},+2),
\label{tripin6^{*}}
\end{eqnarray}
it is the triplet $\Delta_{1}$ introduced in the supersymmetric version of the scheme of 
Gelmini-Roncadelli\footnote{We will call this model as SUSYGR by short.} 
\cite{Rossi:2002zb,Brignole:2003iv,DAmbrosio:2004rko,Rossi:2005ue,Joaquim:2008bm}
\footnote{As we expect the MSUSY331 contain the SUSYGR, and the neutral scalar 
$\sigma^{0}_{1}$ as the scalar $\Delta^{0}_{1}$ generate Majorana mass term for 
the neutrinos \cite{Rodriguez:2022hsj}.}. They are also introduced in chiral superfield and 
anti-chiral supermultiplet is presented in the Tabs.(\ref{lfermionnmssm},\ref{rfermionnmssm}). 

In order to implement supersymmetry, and also at same time to cancel chiral 
anomalies, we must to introduce the following scalars fields \cite{331susy1,mcr}
\begin{eqnarray}
\eta^{\prime} &=& 
\left( 
\begin{array}{c} 
\Phi^{\prime}_{\eta} \\
\eta^{\prime -}_{2}          
\end{array} \right) 
\sim ({\bf1},{\bf \bar{3}},0), \,\
\Phi^{\prime}_{\eta} = \left( 
\begin{array}{c} 
\eta^{\prime 0} \\ 
\eta^{\prime +}_{1}          
\end{array} \right) \sim ({\bf 1},{\bf \bar{2}},+1), \nonumber \\
\rho^{\prime} &=& 
\left( \begin{array}{c} 
\Phi^{\prime}_{\rho} \\
\rho^{\prime --}          
\end{array} 
\right) 
\sim ({\bf1},{\bf \bar{3}},-1), \,\
\Phi^{\prime}_{\rho} = \left( 
\begin{array}{c} 
\rho^{\prime -} \\ 
\rho^{\prime 0}          
\end{array} \right) \sim ({\bf 1},{\bf \bar{2}},+1), \nonumber \\ 
\chi^{\prime} &=& 
\left( \begin{array}{c} 
\Phi^{\prime}_{\chi} \\
\chi^{\prime 0}          
\end{array} \right) 
\sim ({\bf1},{\bf \bar{3}},+1), \,\
\Phi^{\prime}_{\chi} = \left( 
\begin{array}{c} 
\chi^{\prime +} \\ 
\chi^{\prime ++}          
\end{array} \right) \sim ({\bf 1},{\bf \bar{2}},+1), \nonumber \\
S^{\prime}&=& \left( 
\begin{array}{cc} 
T^{\prime} & \frac{\Phi^{\prime}_{S}}{\sqrt{2}} \\ 
\frac{\Phi^{\prime T}_{S}}{\sqrt{2}} & H^{\prime ++}_{2}          
\end{array} \right)
\sim ({\bf 1},{\bf {\bf 6}},0), \,\ 
\Phi^{\prime}_{S}= \left( 
\begin{array}{c} 
h^{\prime -}_{2} \\ 
\sigma^{\prime 0}_{2}          
\end{array} \right) \sim ({\bf 1},{\bf \bar{2}},+1), \nonumber \\ 
T^{\prime}&=&\left( \begin{array}{cc} 
\sigma^{\prime 0}_{1} & \frac{h^{\prime -}_{1}}{\sqrt{2}} \\ 
\frac{h^{\prime }_{1}}{\sqrt{2}} & H^{\prime ++}_{1} \\
\end{array} \right)
\sim ({\bf 1},{\bf \bar{3}},+2). \nonumber \\
\label{3t} 
\end{eqnarray}

In the MSUSY331 we need to introduce the following three vector superfields 
$\hat{V}^{a}_{C}\sim({\bf 8},{\bf 1}, 0)$
\footnote{The gluinos are the superpartner of gluons, and therefore they are in the adjoint representation of $SU(3)$, which is real.}, 
where $a=1,2, \ldots ,8$, $\hat{V}^{a}\sim({\bf 1},{\bf 8}, 0)$, and 
$\hat{V}\sim({\bf 1},{\bf 1}, 0)$. The particle content 
in each vector superfield is presented in the Tab.(\ref{gaugemssm}). 
\begin{table}[h]
\begin{center}
\begin{tabular}{|c|c|c|c|}
\hline 
${\rm{Vector \,\ Superfield}}$ & ${\rm{Gauge \,\ Bosons}}$ & ${\rm{Gaugino}}$ & ${\rm Gauge \,\ constant}$ \\
\hline 
$\hat{V}^{a}_{C}\sim({\bf 8},{\bf 1}, 0)$ & $g^{a}_{m}$ & $\lambda _{C}^{a}$ & $g_{s}$ \\
\hline 
$\hat{V}^{a}\sim({\bf 1},{\bf 8}, 0)$ & $V^{a}_{m}$ & $\lambda^{a}$ & $g$ \\
\hline
$\hat{V}\sim({\bf 1},{\bf 1}, 0)$ & $V_{m}$ & $\lambda$ & $g^{\prime}$ \\
\hline
\end{tabular}
\end{center}
\caption{\small Particle content in the vector superfields in 
MSUSY331, where $a=1,2, \ldots ,8$.}
\label{gaugemssm}
\end{table}

We denote the $SU(3)_{L}$ gauge bosons by $V^{a}_{m}$ ($a=1,2, \ldots ,8$) and since 
\begin{eqnarray}
\left( {\bf 1}, {\bf 8}, 0 \right) \rightarrow 
\left( {\bf 1},{\bf 3}, 0 \right) \oplus
\left( {\bf 1},{\bf 1}, 0 \right) \oplus  
\left( {\bf 1}, {\bf 2}, 3 \right) \oplus 
\left( {\bf 1}, {\bf 2}, -3 \right),
\end{eqnarray} 
we get a triplet, $V^{1}_{m},V^{2}_{m},V^{3}_{m}$, and also a 
singlet $V^{8}_{m}$ both with $Y=0$, plus the doublet of bileptons, 
$\left( V^{4}_{m},V^{5}_{m} \right)$, its hypercharge is 
$Y= 3$ and $\left( V^{6}_{m},V^{7}_{m} \right)$ with $Y=-3$. The charged gauge bosons of 
this model are \cite{331susy1,mcr}
\begin{eqnarray}
W^{ \pm}_{m}(x)&=&-\frac{1}{\sqrt{2}}(V^{1}_{m}(x) 
\mp \imath V^{2}_{m}(x)), \,\
V^{ \pm}_{m}(x)=-\frac{1}{\sqrt{2}}(V^{4}_{m}(x) 
\pm \imath V^{5}_{m}(x)) \nonumber \\
U^{\pm \pm}_{m}(x) &=&- \frac{1}{\sqrt{2}}(V^{6}_{m}(x) 
\pm \imath V^{7}_{m}(x)).
\label{defcarbosons}
\end{eqnarray} 
It was pointed out that process like 
\begin{equation}
e^{-}+e^{-}\rightarrow W^{-}+V^{-},
\end{equation}
violating the unitarity at high energies. One possible solution 
for this problem is the introduction of a doubly charged gauge 
boson $U^{\pm \pm}$ \cite{ppf}. In similar way as in the SM the $Z^{0}$ 
restore the good high energy behavior in the process
\begin{equation}
\bar{\nu} + \nu \rightarrow W^{+}W^{-}.
\end{equation}

The neutral gauge bosons, with $Y=0$, are the photon
\begin{eqnarray}
A_{m}&=& \sin \theta_{W} \left( V^{3}_{m}- \sqrt{3}V^{8}_{m} \right) + 
\sqrt{1-4 \sin^{2} \theta_{W}}V_{m},
\label{photondef}
\end{eqnarray}
where $\theta_{W}$ is the Weinberg angle and we have also two massives neutral bosons defined as
\begin{eqnarray}
Z_{m}&=& \cos \theta_{W}V^{3}_{m}+ \sqrt{3}\tan \theta_{W}\sin \theta_{W}V^{8}_{m}- 
\tan \theta_{W}\sqrt{1-4 \sin^{2} \theta_{W}}V_{m}, \nonumber \\
Z^{\prime}_{m}&=& \frac{1}{\cos \theta_{W}}\left[
\sqrt{1-4 \sin^{2} \theta_{W}}V^{8}_{m}+ \sqrt{3}\sin \theta_{W}V_{m} \right].
\label{Z-ZPdef}
\end{eqnarray}

The Glashow-Iliopoulos-Maiani mechanism, know as GIM-mechanism, for flavor-changing-neutral-currents 
is presented in the following reference \cite{gimm331}. In the MSUSY331, we can write 
\cite{Rodriguez:2022hsj,Rodriguez:2024rvf}
\begin{equation}
\frac{M^{2}_{Z}}{M^{2}_{W}}= \frac{1+4t^{2}}{1+3t^{2}} \equiv \frac{1}{\cos^{2} \theta_{W}}, \,\ 
t \equiv \frac{g^{\prime}}{g},
\label{tinthetawm331}
\end{equation}
we obtain the famous relationship
\begin{equation}
t^{2}= \frac{\sin^{2} \theta_{W}}{1-4 \sin^{2} \theta_{W}},
\label{tinthetaw2}
\end{equation}
To avoid the loss of the model being perturbative, it allows us to write the following inequality \cite{Ng:1992st}
\begin{eqnarray}
\sin^{2}\theta_{W} \left( M_{Z^{\prime}} \right)< \frac{1}{4}.
\label{muesc}
\end{eqnarray} 
When we have equality, we obtain the Landau pole of this model, that is, to obtain this pole, we impose
\begin{eqnarray}
\sin^{2}\theta_{W}(\mu)= \frac{1}{4},
\end{eqnarray}
the Landau scale value of the model without and with SUSY, 
respectively, has the following values \cite{Dias:2004dc}
\begin{eqnarray}
\mu_{\rm M331}&=&5.7 \,\ {\rm TeV}, \nonumber \\
\mu_{\rm MSUSY331}&=&7.8 \,\ {\rm TeV},
\label{dias}
\end{eqnarray}
recently a new analysis appeared in which we obtain the following value 
\cite{Barela:2023oyp}
\begin{eqnarray}
\mu_{\rm M331}&=&8.5 \,\ {\rm TeV}.
\label{barelalimite}
\end{eqnarray}  

This is one of the key reasons for considering the supersymmetric version of this model, as we wish to preserve the model's 
perturbative nature at least up to the TeV scale, which is an energy scale below the GUT scale in which 
\begin{equation}
\sin^{2}\theta_{W}(M_{GUT}) \approx \frac{3}{8}.
\end{equation}
The model predicts five new gauge 
bosons: four charged bileptons $U^{\pm \pm},V^{\pm}$ \cite{Cuypers:1996ia,Montero:1998ve}  and 
one neutral $Z^{\prime}$ \cite{Montero:1998sv}.  

The interaction between the charged bosons with the leptons are given by \cite{mcr,Rodriguez:2010tn}
\begin{eqnarray}
{\cal L}_l^{CC}&=&-\frac{g}{\sqrt{2}}\sum_{l}\left(
\bar{\nu}_{lL}\gamma^{m}V_{\rm PMNS}l_{L}W^{+}_{m}+ 
\bar{l}^{c}_{L}\gamma^{m}U_{V}\nu_{lL} V^{+}_{m}+
\bar{l}^{c}_{L}\gamma^{m} U_{U}l_{L}U^{++}_{m}+hc \right). \nonumber \\
\label{lf}
\end{eqnarray}
The gauge bosons $V^{\pm}$ and $U^{\pm \pm}$ are known as bileptons \cite{Cuypers:1996ia}. The 
$V_{\rm PMNS}$ is the Pontecorvo-Maki-Nakagawa-Sakata mixing matrix as defined in 
Eq.(\ref{pmnsdef}). There are new mixing matrices given by $U_{V}$ and $U_{U}$. The same 
interactions with the quarks and gauge bosons are
\begin{eqnarray}
{\cal L}_q^{CC}& = & -\frac{g}{2\sqrt{2}}\left[\overline{U}\gamma^m(1 - 
\gamma_5)V_{\rm CKM}DW^+_m  + \overline{U}\gamma^m(1 - \gamma_5)\zeta{\cal 
JV}_m \right. \nonumber \\ &+& \left. 
\overline{D}\gamma^m(1 - \gamma_5)\xi{\cal JU}_m\right] + hc, 
\nonumber \\
\label{lq}
\end{eqnarray}
where we have defined the mass eigenstates in the following way
\begin{eqnarray}
U &=& \left(\begin{array}{c} u \\ c \\ t
\end{array}\right), \quad
D =  \left(\begin{array}{c}
         d \\ s \\ b
\end{array}\right), \quad
{\cal V}_m = \left(\begin{array}{c}
         V^+_m \\ U^{--}_m \\ U^{--}_m\end{array}\right), \nonumber \\
{\cal U}_m & = & \left(\begin{array}{c}
         U^{--}_m \\ V^+_m \\ V^+_m
\end{array}\right), \,\ 
{\cal J}=  \left(\begin{array}{c}
        J_{1} \equiv j_{1} \\ J_{2} \equiv j_{2} \\ J_{3} \equiv J
\end{array}\right).  
\label{maest}
\end{eqnarray}
The $V_{\rm CKM}$ is the usual Cabibbo-Kobayashi-Maskawa (CKM) mixing matrix, 
defined in Eq.(\ref{lq}),  and $\xi$ 
and $\zeta$ are mixing matrices containing new unknown mixing parameters due to 
the presence of the exotic quarks.

We can define the charged gauginos in analogy with the gauge 
bosons, defined in our 
Eqs.(\ref{defcarbosons},\ref{photondef},\ref{Z-ZPdef}), in the following way \cite{331susy1,mcr}
\begin{eqnarray}
\lambda_{W}^{\pm}(x)&=& -\frac{1}{\sqrt{2}}(\lambda^{1}_{A}(x) \mp \imath 
\lambda^{2}_{A}(x)), \nonumber \\
\lambda_{V}^{\pm}(x)&=&-\frac{1}{\sqrt{2}}(\lambda^{4}_{A}(x) \pm \imath 
\lambda^{5}_{A}(x)),\nonumber \\
\lambda_{U}^{\pm \pm}(x) &=&-\frac{1}{\sqrt{2}}(\lambda^{6}_{A}(x) \pm \imath 
\lambda^{7}_{A}(x)).
\label{defgauginos}
\end{eqnarray}

The Lagrangian of this model is presented in the references \cite{331susy1,mcr}. In this model,
we can define as in the MSSM the following $R$-Parity 
\cite{Rodriguez:2024rvf,Rodriguez:2010tn}
\begin{equation}
R=(-1)^{2S+3{\cal F}},
\label{rparitymsusy331}
\end{equation} 
where $S$ is the spin and we define the following new quantum number \cite{Rodriguez:2010tn}
\begin{equation}
{\cal F}\equiv B+L
\end{equation}  
where $B$ is the baryon number while $L$ is the total lepton number. The ${\cal F}$ number attribution is
\begin{equation}
\begin{array}{c}
{\cal F}(U^{--})={\cal F}(V^{-}) = - {\cal F}(J)= {\cal F}(j_{1,2})=
{\cal F}(\rho^{--})  \\  
= {\cal F}(\chi^{--}) ={\cal F}(\chi^{-}) = 
{\cal F}(\eta^{-}_{2})={\cal F}(\sigma_{1}^{0})=2.
\end{array}
\label{efe}
\end{equation}  
Choosing the following R-charges 
\begin{eqnarray}
n_{\eta}&=&n_{\rho^{\prime}}=n_{S}=-1, \,\
n_{\rho}=n_{\eta^{\prime}}=n_{S^{\prime}}=1, \,\
n_{\chi}=n_{\chi^{\prime}}=0, \nonumber \\
n_{L}&=&n_{Q_{i}}=n_{d_{i}}=1/2, \,\
n_{J_{i}}=-1/2, \,\ n_{u}=-3/2,
\label{rdiscsusy331} 
\end{eqnarray}
the superpotential of our model that conseves the $R$-Parity, defined in 
Eq.(\ref{rparitymsusy331}), is given by
\begin{eqnarray} 
W^{MSUSY331}_{RPC}&=&\mu_{ \eta} \hat{ \eta} \hat{ \eta}^{\prime}+
\mu_{ \rho} \hat{ \rho} \hat{ \rho}^{\prime}+
\mu_{ \chi} \hat{ \chi} \hat{ \chi}^{\prime}+ 
+
\mu_{S} Tr[(\hat{S} \hat{S}^{\prime})]+
\lambda_{2ij} \epsilon \hat{L}_{iL} \hat{L}_{jL} \hat{ \eta} \nonumber \\
&+&
\lambda_{3ij} 
(\hat{L}_{i}\hat{S}\hat{L}_{j}) +
f_{1} \epsilon \hat{ \rho} \hat{ \chi} \hat{ \eta}+
f_{2} \epsilon \hat{ \chi} \hat{ \rho} \hat{S}+ 
f^{\prime}_{1}\epsilon \hat{ \rho}^{\prime}\hat{ \chi}^{\prime}
\hat{ \eta}^{\prime} +
f^{\prime}_{2} \epsilon \hat{ \chi}^{\prime} \hat{ \rho}^{\prime} 
\hat{S}^{\prime} \nonumber \\ &+&
\kappa_{1\alpha i} \left( \hat{Q}_{\alpha L} \hat{\eta} \right) \hat{d}^{c}_{iL} 
+
\kappa_{2\alpha i} \left( \hat{Q}_{\alpha L} \hat{\rho} \right) \hat{u}^{c}_{iL} +
\kappa_{3\alpha \beta} \left( \hat{Q}_{\alpha L} \hat{\chi} \right) \hat{j}^{c}_{\beta L} 
\nonumber \\ &+&
\kappa_{4i} \left( \hat{Q}_{3 L} \hat{\eta}^{\prime} \right) \hat{u}^{c}_{jL}+
\kappa_{5i} \left( \hat{Q}_{3L} \hat{\rho}^{\prime} \right) \hat{d}^{c}_{iL}+
\kappa_{6 \alpha \beta} \left( \hat{Q}_{\alpha L} \hat{\chi} \right) \hat{j}^{c}_{\beta L} \nonumber \\ &+&
\kappa_{7} \left( \hat{Q}_{3L} \hat{\chi}^{\prime} \right) \hat{J}^{c}_{L}. \nonumber \\
\label{supmsusy331rparitycons}
\end{eqnarray}
The coupling $\kappa_{1\alpha i}, \kappa_{2\alpha i}$ are symmetric 
in the indices $\alpha$ and $i$; the same is hold for $\kappa_{6 \alpha \beta}$, it is symmetric 
in the indices $\alpha$ and $\beta$.

\section{The masses of Fermions}

We have already presented a preliminar analyses of it without the sextet ($S^{\prime}$) and the anti-sextet ($S$) 
\cite{Rodriguez:2010tn}. For our numercial analyses we will choose the follwing vacuum expectation values (VEV) in (GeV) \cite{331susy1,Rodriguez:2022hsj,Rodriguez:2024rvf}
\begin{eqnarray}
v_{\eta}=20, \,\ v_{\sigma_{2}}=10, \,\ 
v_{\eta^{\prime}}=v_{\rho^{\prime}}=1, \,\ 
v_{\rho}=245.198.
\label{vcp}
\end{eqnarray}

\subsection{Masses for the usuais Quarks.}
\label{massquarksdown}

The Yukawa terms for the up quarks are
\begin{equation}
- \left[ \kappa _{2 \alpha i} 
\left( Q_{\alpha }\rho \right) \hat{u}^{c}_{i}+ \kappa_{4i}
\left( Q_{3}\eta^{\prime} \right) \hat{u}^{c}_{i} \right] +hc,
\end{equation}
while for down quarks we have
\begin{equation}
- \left[ \kappa _{1 \alpha i} 
\left( Q_{\alpha }\eta \right) \hat{d}^{c}_{i}+ \kappa_{5i}
\left( Q_{3}\rho^{\prime} \right) \hat{d}^{c}_{i} \right] +hc.
\end{equation}

We can define the general mass term for any quark in the following way \cite{Rodriguez:2010tn}
\begin{equation}
- \left[ \left( \frac{1}{2} \right)
\Psi _{q}^{\pm T}Y_{q}^{ \pm}\Psi _{q}^{\pm }+ hc \right]
\label{generalmasstermquarks}
\end{equation}
where $Y_{q}^\pm$ is given by 
\begin{equation}
Y_{q}^{ \pm}=
\left( 
\begin{array}{cc}
0 & X_{q}^{T} \\ 
X_{q} & 0
\end{array}
\right),
\label{ypm}
\end{equation}
with
\begin{eqnarray}
X_{u}&=&\frac{1}{\sqrt2}\,\left(\begin{array}{ccc}
\kappa_{211}v_{\rho} & \kappa_{212}v_{\rho} & \kappa_{213}v_{\rho} \\
\kappa_{221}v_{\rho} & \kappa_{222}v_{\rho} & \kappa_{223} v_{\rho} \\
\kappa_{41}v_{\eta^{\prime}} & \kappa_{42}v_{\eta^{\prime}} &
\kappa_{43}v_{\eta^{\prime}}\\ 
\end{array}\right), \nonumber \\
X_{d}&=&\frac{1}{\sqrt2}\,\left(\begin{array}{ccc}
\kappa_{111}v_{\eta} & \kappa_{112}v_{\eta} & \kappa_{113}v_{\eta} \\
\kappa_{121}v_{\eta} & \kappa_{122}v_{\eta} & \kappa_{123}v_{\eta} \\
\kappa_{51}v_{\rho^{\prime}} & \kappa_{52}v_{\rho^{\prime}} &
\kappa_{53}v_{\rho^{\prime}}\\ 
\end{array}\right).
\label{clmm1}
\end{eqnarray}

The q-quarks mass matrix is diagonalized using two rotation matrices, 
$H$ and $I$, defined by
\begin{equation}
M_{q}^{2}=H \left( X_{q}^{T}X_{q} \right) H^{-1}=
I^{*} \left( X_{q}X_{q}^{T} \right) \left( I^{*} \right)^{-1}.
\label{m2j}
\end{equation}

We want to emphasize the following facts
\begin{itemize}
\item[a-)] A familly of quarks has different transformation properties than 
the others two families, see our Eqs.(\ref{q1l},\ref{q23l});
\item[b-)] We need two different VEV to generate masses for the up quarks, 
$\eta^{\prime}$ and $\rho$, and down quarks, $\rho^{\prime}$ and $\eta$, 
as discussed briefly in our reference \cite{331susy1,Rodriguez:2022hsj};
\end{itemize} 
with this we can hope to have a simple explanation of the why the quarks of 
the first family are much more ligther than the quarks of the other two 
families.

We will impose the following discrete symmetry \cite{cmmc,cmmc1}
\begin{equation}
\hat{u}^{c}_{3} \rightarrow - \hat{u}^{c}_{3}, \,\ 
\hat{d}^{c}_{3} \rightarrow - \hat{d}^{c}_{3}, 
\end{equation}
all others superfields are even, except $\hat{Q}_{3}$. For these quarks, there are two possibilities 
\begin{itemize}
\item[1-)] $\hat{Q}_{3}$ is even and the masses for the quarks ``up" and ``down" are given by \cite{cmmc}
\begin{eqnarray}
m_{u}  & \propto& \frac{\alpha_{s}\sin(2\theta_{\tilde{u}})}{\pi}m_{\tilde{g}}
\left[  \frac{M_{\tilde{u_{1}}}^{2}}{M_{\tilde{u_{1}}}^{2}-m_{\tilde{g}}^{2}}
\ln\left(  \frac{M_{\tilde{u_{1}}}^{2}}{m_{\tilde{g}}^{2}}\right)  \right.
\nonumber\\
& -& \left.  \frac{M_{\tilde{u_{2}}}^{2}}{M_{\tilde{u_{2}}}^{2}-m_{\tilde{g}}^{2}}
\ln\left(  \frac{M_{\tilde{u_{2}}}^{2}}{m_{\tilde{g}}^{2}}\right)
\right]  \,\ ,\nonumber\\
m_{d}  & \propto& \frac{\alpha_{s}\sin(2\theta_{\tilde{d}})}{\pi}m_{\tilde{g}}
\left[  \frac{M_{\tilde{d_{1}}}^{2}}{M_{\tilde{d_{1}}}^{2}-m_{\tilde{g}}^{2}}
\ln\left(  \frac{M_{\tilde{d_{1}}}^{2}}{m_{\tilde{g}}^{2}}\right)  \right.
\nonumber\\
& -& \left.  \frac{M_{\tilde{d_{2}}}^{2}}{M_{\tilde{d_{2}}}^{2}-m_{\tilde{g}}^{2}}
\ln\left(  \frac{M_{\tilde{d_{2}}}^{2}}{m_{\tilde{g}}^{2}}\right)
\right];
\label{udeeoneloop}
\end{eqnarray}
where $m_{\tilde{g}},m_{\tilde{u}},m_{\tilde{d}}$ and $m_{\tilde{s}}$ are the masses of gluinos and squarks.  If we take
\begin{eqnarray}
m_{\tilde{g}} &\approx& 100 \,\ {\mbox GeV}; \nonumber \\
\sin \left( 2 \theta_{\tilde{u}} \right)&=& \sin \left( 2 \theta_{\tilde{d}} \right) \approx 10^{-3},
\end{eqnarray}
we obtain the correct values for the masses of those quarks.
\item[2-)] $\hat{Q}_{3}$ is odd and under these hypotheses, the following two interesting possibilities exist
\item[2a-)] Light Quarks
\begin{eqnarray}
m_{u}&=& \kappa_{43} \frac{v_{\eta^{\prime}}}{\sqrt{2}} \,\ \rightarrow
\kappa_{43}= \frac{\sqrt{2}m_{u}}{v_{\eta^{\prime}}}=0.00707; \nonumber \\
m_{d}&=& \kappa_{53} \frac{v_{\rho^{\prime}}}{\sqrt{2}} \,\ \rightarrow
\kappa_{53}= \frac{\sqrt{2}m_{d}}{v_{\rho^{\prime}}}=0.0127;
\label{quarksleve}
\end{eqnarray}
We take $v_{\eta^{\prime}}=v_{\rho^{\prime}}$ it is in agreement with 
the results we get to their masses using baryon magnetic moments \cite{donoghue}
\begin{equation}
m_{u}\sim m_{d}\sim 320 \,\ {\mbox MeV}.
\end{equation}
\item[2b-)] Heavy Quarks
\begin{eqnarray}
\delta_{u}&=& \kappa^{2}_{211}+2 \kappa^{2}_{212}+ \kappa^{2}_{222}, \,\ 
\Delta_{u}= \delta_{u}^{2}-4 
\left( \kappa_{211}\kappa_{222}- \kappa^{2}_{212} \right)^{2}, \nonumber \\
M^{2}_{c}&=& \frac{v^{2}_{\rho}}{4}\left[ \delta_{u} - \sqrt{\Delta_{u}} \right], \,\
M^{2}_{t}= \frac{v^{2}_{\rho}}{4}\left[ \delta_{u} + \sqrt{\Delta_{u}} \right], \nonumber  \\
\delta_{d}&=& \kappa^{2}_{111}+2 \kappa^{2}_{112}+ \kappa^{2}_{122}, \,\
\Delta_{d}= \delta_{d}^{2}-4 
\left( \kappa_{111}\kappa_{122}- \kappa^{2}_{112} \right)^{2}, \nonumber \\
M^{2}_{s}&=& \frac{v^{2}_{\eta}}{4}\left[ \delta_{d} - \sqrt{\Delta_{d}} \right], \,\
M^{2}_{b}= \frac{v^{2}_{\eta}}{4}\left[ \delta_{d} + \sqrt{\Delta_{d}} \right].
\label{quarkspesados}
\end{eqnarray}
\end{itemize}
We can also write the following relations
\begin{eqnarray}
M^{2}_{c}&+&M^{2}_{t}= \frac{v^{2}_{\rho}}{2} \delta_{u} , \,\  
M^{2}_{c}\times M^{2}_{t}=\frac{v^{4}_{\rho}}{4} 
\left( \kappa_{211}\kappa_{222}- \kappa^{2}_{212} \right), \nonumber \\
M^{2}_{s}&+&M^{2}_{b}= \frac{v^{2}_{\eta}}{2} \delta_{d} , \,\  
M^{2}_{s}\times M^{2}_{b}=\frac{v^{4}_{\eta}}{4} 
\left( \kappa_{111}\kappa_{122}- \kappa^{2}_{112} \right).
\label{quarkspesados}
\end{eqnarray}
We have choose $v_{\rho}> v_{\eta}$ it is natural to explain the result that the quarks 
$t$ and $s$ are more massive than quarks $c$ and $b$. We can fit the masses of those quarks in 
terms of these parameters $\kappa_{211},\kappa_{212},\kappa_{222}$, in terms of 
$M_{c}$ and $M_{t}$, by means of the following relations
\begin{eqnarray}
\kappa_{222}&=&\frac{2(M_{c}+M_{t})}{v^{2}_{\rho}}- \kappa_{211}, \nonumber \\
\kappa^{2}_{212}&=&\kappa_{211}\kappa_{212}- \frac{2M_{c}M_{t}}{v^{2}_{\rho}},
\label{kkpossiveis}
\end{eqnarray}  
and it is shown in Figs.(). The same for $M_{s}$ and $M_{b}$ are 
in Figs.().

The  eigenvector are
\begin{eqnarray}
E_{L}^{u}&=&\left(
\begin{array}
[c]{ccc}
1 & 0 & 0 \\
0 & \cos\theta_{u} & \sin\theta_{u} \\
0 &-\sin\theta_{u} & \cos\theta_{u} 
\end{array}
\right)  , \,\
E_{L}^{d}= \left(
\begin{array}[c]{ccc}
1 & 0 & 0 \\
0 & \cos\theta_{d} & \sin\theta_{d} \\
0 &-\sin\theta_{d} & \cos\theta_{d}
\end{array}
\right)  \,\ ,
\label{autovetor2}
\end{eqnarray}
Then, with 
we can get an expression to the CKM matrix, see our Eq.(\ref{lq}), as follows
\footnote{To understand this definition remember the discussion presented before our Eq.(\ref{datackm}).}:
\begin{equation}
V_{CKM}= \left(E_{L}^{u} \right)^{\dagger}E_{L}^{d}=\left(
\begin{array}[c]{ccc}
1 & 0 & 0 \\
0 & \cos \left( \theta_{u}- \theta_{d} \right) & \sin \left( \theta_{u}- \theta_{d} \right)  \\
0 &-\sin \left( \theta_{u}- \theta_{d} \right)  & \cos \left( \theta_{u}- \theta_{d} \right) 
\end{array}
\right)  \,\ .
\label{ckmz2df}
\end{equation}
We can conclude the $V_{CKM}$, defined in Eq.(\ref{ckmz2df}), gives a hint about the mixing 
angles given in Eq.(\ref{datackm}).

\subsection{Masses for Exotic Quarks}
\label{massquarksexotics}
For the J-quark type. There are interactions like
\begin{equation}
- \left[ \kappa_{7} \left( Q_{3} \chi^{\prime }J^{c}+
\bar{Q}_{3} \overline{\chi^{\prime }} \bar{J}^{c} \right) \right],
\end{equation}
which imply one diagonalized state with the following mass
\begin{equation}
M_{J}^{mass}=- \frac{\kappa_{7}}{\sqrt{2}} v_{\chi^{\prime}} 
\left( JJ^{c}+ \bar{J}\bar{J^{c}} \right).
\label{Jmass}
\end{equation}

We shown in Fig. the values of $M_{J}$ as function of 
$v_{\chi^{\prime}}$. We can see if we consider 
\begin{equation}
0.7 \leq \kappa_{7} \leq 1.0, 
\end{equation}
we get 
\begin{equation}
M_{J}>4.0 {\mbox TeV},
\end{equation}
for $v_{\chi^{\prime}} \approx 8.0$ TeV.

The j-quarks mass matrix is diagonalized using two rotation matrices, 
$H$ and $I$, defined by
\begin{eqnarray}
j_{\alpha }^{+}=H_{\alpha \beta }\psi _{j_{\beta }}^{+}, \,\
j_{\alpha }^{-}=I_{\alpha \beta }\psi _{j_{\beta }}^{-}, \,\  \alpha ,\beta
=1,2.
\label{2scj} 
\end{eqnarray}
We can write the diagonal matrix ($H$ and $I$ are unitary) in the following way
\begin{eqnarray}
M_{j}&=&I^{*}X_{j}H^{-1}, \nonumber \\
M_{j}^{2}&=&H \left( X_{j}^{T}X_{j} \right) H^{-1}=
I^{*} \left( X_{j}X_{j}^{T} \right) \left( I^{*} \right)^{-1}.
\label{m2j}
\end{eqnarray}
The masses of physical $j$ are
\begin{eqnarray}
\delta_{j}&=& \kappa^{2}_{311}+2 \kappa^{2}_{312}+ \kappa^{2}_{322}, \,\
\Delta_{j}= \delta^{2}_{j}-4 
\left( \kappa_{311}\kappa_{322}- \kappa^{2}_{312} \right)^{2}, \nonumber \\
M^{2}_{j_{1}}&=& \frac{v^{2}_{\chi}}{4}\left[ \delta_{j} - \sqrt{\Delta}_{j} \right], \,\
M^{2}_{j_{2}}= \frac{v^{2}_{\chi}}{4}\left[ \delta_{j} + \sqrt{\Delta}_{j} \right], 
\end{eqnarray}
from these relations, we can write
\begin{eqnarray}
M^{2}_{j_{1}}&+&M^{2}_{j_{2}}= \frac{v^{2}_{\chi}}{2} \delta_{j}, \nonumber \\  
M^{2}_{j_{1}}&\times& M^{2}_{j_{2}}=\frac{v^{4}_{\chi}}{4} 
\left( \kappa_{311}\kappa_{322}- \kappa^{2}_{312} \right).
\label{j1j2exact}
\end{eqnarray}

Remember the mass $M_{J}$, defined in Eq.(\ref{Jmass}), is proportional 
to $v_{\chi^{\prime}}$ and $M_{j}$ are proportional to $v_{\chi}$ and as 
in general we consider $v_{\chi^{\prime}}>v_{\chi}$ it will imply 
$M_{J}>M_{j}$. We shown in Fig.() the behaviour of $M_{J}$, we assume 
in this case $v_{\chi^{\prime}}\approx v_{\chi}$, and $M_{j_{1}}$ as function of 
$v_{\chi}$.

We shown in our Fig.(), the values of $M_{j_{1}}$ and $M_{j_{2}}$ 
as function of $v_{\chi}$ when $\kappa_{311}=0.5$ and $\kappa_{312}=0.6$.

We shown in Fig.() the values of $M_{j_{1}}$ as function of $v_{\chi}$. We can see 
if we consider 
\begin{equation}
\kappa_{322} \leq 0.5, 
\end{equation}
for $v_{\chi} \approx 4$ TeV. 
is easy explain

\section{Masses for Charged Leptons}
\label{masscharlep}

Let us first considered the charged leptons and its Yukawa term is given by the following expression
\begin{eqnarray}
- \left[ \lambda_{2ij} \epsilon_{ijk}L_{i}L_{j}\eta_{k}+ 
\lambda_{3ij} L_{i}S_{ij}L_{j}+ hc \right].
\label{newint}
\end{eqnarray}
This term will generate the following mass matrix for the charged leptons \cite{331susy1}
\begin{equation}
M^{l}_{ij}= \frac{\lambda_{2ij}}{\sqrt{2}}v_{\eta} +
\frac{\lambda_{3ij}}{\sqrt{2}} v_{\sigma^{0}_{2}}.
\label{charleptonmass}
\end{equation} 
where $v_{\eta}$ is the VEV of the $\langle \eta^{0} \rangle$, the triplet $\eta$, while $v_{\sigma^{0}_{2}}$ is the VEV of
the  $\langle\sigma^{0}_{2}\rangle$, the anti-sextet $S$. 

We can perform the same analyses done at quarks sector. But now, we will present the most 
simple analyse for this sector. It is, we can assume
\begin{eqnarray}
\lambda_{212}&=&\lambda_{213}=\lambda_{223}=0, \nonumber \\
\lambda_{312}&=&\lambda_{313}=\lambda_{323}=0,
\label{hip1}
\end{eqnarray}
then we can write
\begin{eqnarray}
\lambda_{311}&\equiv& \frac{m_{e}}{\sqrt{2} v_{\sigma_{2}}}=0.000025, \nonumber \\
\lambda_{322}&\equiv& \frac{m_{\mu}}{\sqrt{2} v_{\sigma_{2}}}=0.00525, \nonumber \\
\lambda_{333}&\equiv& \frac{m_{\tau}}{\sqrt{2} v_{\sigma_{2}}}=0.0885.
\label{hip2}
\end{eqnarray}
Therefore the Eqs.(\ref{charleptonmass},\ref{hip2}), can explain 
the masses of charged leptons.

\section{Masses for Neutrinos}
\label{massneutrinos}

If $\langle \sigma^{0}_{1}\rangle \neq 0$ neutrinos get the following  Majorana mass term 
\begin{equation}
M^{\nu}_{ij}= \frac{\lambda_{3ij}}{\sqrt{2}}v_{\sigma^{0}_{1}}.
\label{neutrinomass}
\end{equation}
However, if we consider $v_{\eta}=0$ then $M^{l} \propto M^{\nu}$ and it would 
imply that the Pontecorvo-Maki-Nakagawa-Sakata (PMNS) matrix obey
\begin{equation}
U_{PMNS}=I_{3 \times 3}.
\end{equation} 
As happen in the SM, we can fit Eq.(\ref{pmnsdef}), if we suppose
\begin{itemize}
\item The mixing $\theta_{23}$ is consistent with maximal mixing;
\item The mixing $\theta_{12}$ is large but not maxima;
\item The CHOOZ (in France) results indicate a tiny value for the 
mixing angle $\theta_{13}$.
\end{itemize}

We can consider two scenarios\footnote{Here we have 
omitted the Majorana phases because they do not lead 
to observable effects in oscillations.}:
\begin{itemize}
\item ``bi-large'' mixing, where the mixing 
parameters are $\theta_{23} = \pi/4$ and $\theta_{13}=0$
\begin{equation}
U_{BL} = 
\left(
\begin{array}{ccc}
\cos\theta_{12} & \sin\theta_{12} & 0 \\
-\frac{\sin\theta_{12}}{\sqrt{2}} & 
\frac{\cos\theta_{12}}{\sqrt{2}} & 
\frac{1}{\sqrt{2}} \\
-\frac{\sin\theta_{12}}{\sqrt{2}} & 
\frac{\cos\theta_{12}}{\sqrt{2}} & 
-\frac{1}{\sqrt{2}}
\end{array}
\right) 
\label{eq1:osc_BiLarge}
\end{equation}
\item ``tribimaximal'' mixing, where the mixing 
parameters are $\theta_{12}$ is very well approximated by the relation: $\sin^2 \theta_{12} = 1/3$ 
and the Eq.(\ref{eq1:osc_BiLarge}) can be written as
\begin{equation}
U_{TB} = 
\left(
\begin{array}{ccc}
\sqrt{\frac{2}{3}} & \frac{1}{\sqrt{3}} & 0 \\
-\frac{1}{\sqrt{6}} & 
\frac{1}{\sqrt{3}} & \frac{1}{\sqrt{2}} \\
-\frac{1}{\sqrt{6}} & 
\frac{1}{\sqrt{3}} & -\frac{1}{\sqrt{2}}
\end{array}
\right). 
\label{eq1:osc_TriBi}
\end{equation} 
\end{itemize}

\section{Mixing between Gauginos and Higgsinos}

As happen in the Minimal Supersymmetric Standard Model, more known as MSSM, the 
gauginos and higgsinos can mix and we get charginos and neutralinos. We will get their mass spectrum in the next sub-sections.

\subsection{Double Charged Charginos}

We have already analyzed this sector, and the mass matrix is given by \cite{mcr}
\begin{equation}
T= \left( \begin{array}{ccccc}
-m_{\lambda}& -gu& gw^{\prime}& \frac{gz}{\sqrt{2}}& - \frac{gz^{\prime}}{\sqrt{2}} \\
gu^{\prime}& \frac{\mu_{\rho}}{2}&- 
\left( \frac{f_{1}^{\prime}v^{\prime}}{3}- \sqrt{2} \frac{f_{2}^{\prime}}{3}z^{\prime} \right)& 
0& \frac{f_{2}^{\prime}}{3}w^{\prime} \\
-gw&- \left( \frac{f_{1}v}{3}- \sqrt{2} \frac{f_{2}}{3}z \right)& \frac{\mu_{\chi}}{2}& 
\frac{f_{2}}{3}u& 0 \\
- \frac{gz^{\prime}}{\sqrt{2}}& 0& \frac{f_{2}^{\prime}}{3}u^{\prime}& \frac{\mu_{S}}{2}& 0 \\
\frac{gz}{\sqrt{2}}& \frac{f_{2}}{3}w& 0& 0& \frac{\mu_{S}}{2}
\end{array}
\right).
\end{equation}
We will consider the following VEV in GeV given by Eq.(\ref{vcp}), and we will also to 
choose the following values for our parameters in GeV
\begin{eqnarray}
\mu_{\eta}&=&300, \,\  \mu_{\rho}=500, \mu_{\chi}=700, \nonumber \\
m_{\lambda}&=&3000, \,\ \mu_{S}=500, 
\label{pdmsusy331rc}
\end{eqnarray}
together with
\begin{eqnarray}
g&=&0.65, \,\ f_{1}=0.151, \,\ f^{\prime}_{1}=1.5, \,\ f_{2}=1.0, 
f^{\prime}_{2}=1.0,
\label{pamsusy331rc}
\end{eqnarray}

Our results are shown in Figs.(), where we can note, there are 
the following superior limit in (GeV) 
\begin{itemize}
\item For the lighest double chargino $\tilde{\chi}^{\pm \pm}_{1}$ see also 
Figs.()
\begin{equation}
m \leq 300 \,\ {\mbox GeV}
\end{equation}
\item For $\tilde{\chi}^{\pm \pm}_{2}$
\begin{equation}
m \leq 600 \,\ {\mbox GeV}.
\end{equation}
\end{itemize}

\subsection{Charginos}

In this case we get the following mass matrix \cite{mcr}
\begin{equation}
X= \left( \begin{array}{cccccccc}
- m_{\lambda}& 0& gv^{\prime}& 0&- gu& 0&- \frac{gz^{\prime}}{2}& 0 \\
0&- m_{\lambda}& 0&- gv& 0& gw^{\prime}& 0&- \frac{gz}{2}  \\
-gv& 0& \frac{\mu_{\eta}}{2}& 0&- \frac{f_{1}w}{3}& 0& 0& 0 \\
0& gv^{\prime}& 0& \frac{\mu_{\eta}}{2}& 0& \frac{f^{\prime}_{1}u^{\prime}}{3}& 0& 0 \\
gu^{\prime}& 0&- \frac{f^{\prime}_{1}w^{\prime}}{3}& 0& \frac{\mu_{\rho}}{2}& 0& 0& 0 \\
0&- gw& 0& \frac{f_{1}u}{3}& 0& \frac{\mu_{\chi}}{2}& 0& 0 \\
- \frac{gz}{2}& 0& 0& 0& \frac{f_{3}w}{3 \sqrt{2}}& 0& \frac{\mu_{S}}{2}& 0 \\
0 & \frac{gz^{ \prime}}{2}& 0& 0& 0& \frac{f^{ \prime}_{3}u^{ \prime}}{3 \sqrt{2}}& 0& 
\frac{\mu_{S}}{2}
\end{array}
\right).
\end{equation}

Using Eqs.(\ref{vcp},\ref{pdmsusy331rc},\ref{pamsusy331rc}), we get the result presented in 
Fig.(), where we can note, there are 
the following superior limit in (GeV) 
\begin{itemize}
\item For the lighest chargino $\tilde{\chi}^{\pm}_{1}$
\begin{equation}
m \approx 100 \,\ {\mbox GeV},
\end{equation}
\item For $\tilde{\chi}^{\pm}_{2}$
\begin{equation}
m \leq 300 \,\ {\mbox GeV},
\end{equation}
\end{itemize}
these results are in agreement with the experimental limits \cite{GAMBIT:2018gjo}, see also our Eq.(\ref{limitesnoMSSM}).

\subsection{Neutralinos}

The mass matrix for neutralinos is \cite{mcr} 
\begin{eqnarray}
Y^{0}&=& \left( \begin{array}{cc}
{\cal M}_{\tilde{G}\tilde{G}} & {\cal M}_{\tilde{G}\tilde{H}} \\
{\cal M}^{T}_{\tilde{G}\tilde{H}} & {\cal M}_{\tilde{H}\tilde{H}}
\end{array}
\right), \nonumber \\
{\cal M}_{\tilde{G}\tilde{G}}&=&- \left( \begin{array}{ccc}
m_{\lambda} & 0 & 0 \\
0 & m_{\lambda} & 0 \\
0 & 0 & m^{\prime}
\end{array}
\right), \nonumber \\ 
{\cal M}_{\tilde{G}\tilde{H}}&=& \left( \begin{array}{cccccccccc}
- \frac{gv}{\sqrt{2}}& \frac{gv^{\prime}}{\sqrt{2}}& \frac{gu}{\sqrt{2}}&- 
\frac{gu^{\prime}}{\sqrt{2}}& 0& 0& 0& 0& \frac{gz}{2 \sqrt{2}}&- 
\frac{gz^{\prime}}{2 \sqrt{2}} \\
- \frac{gv}{\sqrt{6}}& \frac{gv^{\prime}}{\sqrt{6}}&- 
 \frac{gu}{\sqrt{6}}& \frac{gu^{\prime}}{\sqrt{6}}& 
\frac{2}{\sqrt{6}} gw&- \frac{2}{\sqrt{6}}gw^{\prime}& 0& 0& \frac{gz}{2 \sqrt{6}}&- 
\frac{gz^{\prime}}{2 \sqrt{6}}\\
0& 0&- \frac{g^{\prime}u}{\sqrt{2}}& 
\frac{g^{\prime}u^{\prime}}{\sqrt{2}}&  
\frac{g^{\prime}w}{\sqrt{2}}&- \frac{g^{\prime}w^{\prime}}{\sqrt{2}}& 0 & 0& 0& 0 \\
- \frac{gv}{\sqrt{2}}&- \frac{gv}{\sqrt{6}}& 0& 0& \frac{\mu_{\eta}}{2}& 
\frac{f_{1}w}{3}& 0&- \frac{f_{1}u}{3}& 0& 0 
\end{array}
\right), \nonumber \\
{\cal M}_{\tilde{H}\tilde{H}}&=& \left( \begin{array}{cccccccccc}
0& \frac{\mu_{\eta}}{2}& 
\frac{f_{1}w}{3}& 0&- \frac{f_{1}u}{3}& 0& 0& 0& 0& 0 \\
\frac{\mu_{\eta}}{2}& 
0& 0& \frac{f^{\prime}_{1}w^{\prime}}{3}& 0&- \frac{f^{\prime}_{1}u^{\prime}}{3}& 
0& 0& 0& 0 \\
\frac{f_{1}w}{3}& 0& 0& \frac{ \mu_{\rho}}{2}& A 
& 0& 0& 0& \frac{f_{3}w}{3 \sqrt{2}}& 0 \\
0& \frac{f^{\prime}_{1}w^{\prime}}{3}& 
\frac{ \mu_{\rho}}{2}& 0& 0& B& 0& 0& 0& \frac{f^{\prime}_{3}w^{\prime}}{3 \sqrt{2}} \\
- \frac{f_{1}u}{3}& 0& 
A& 0& 0& \frac{ \mu_{\chi}}{2}& 0& 0& \frac{f_{3}u}{3 \sqrt{2}}& 0 \\
0&- \frac{f^{\prime}_{1}u^{\prime}}{3}& 0& B& \frac{ \mu_{\chi}}{2}& 0& 0& 
0& 0& \frac{f^{\prime}_{3}u^{\prime}}{3 \sqrt{2}} \\
0& 0& 0& 0& 0& 0& 0& \frac{ \mu_{S}}{2}& 0&  0 \\
0& 0& 0& 0& 0& 0& 0& 0& 0& \frac{ \mu_{S}}{2} \\
\frac{gz^{\prime}}{2 \sqrt{2}}& \frac{gz}{2 \sqrt{6}}& 0& 0& 0& \frac{f_{3}w}{3 \sqrt{2}}& 0& 
\frac{f_{3}u}{3 \sqrt{2}} & 0 & 0  \\
0& 0& 0& 0& 0& 0& 0& 0& \frac{ \mu_{S}}{2} & 0 \\
\frac{f^{\prime}_{3}w^{\prime}}{3 \sqrt{2}}& 0& \frac{f^{\prime}_{3}u^{\prime}}{3 \sqrt{2}}& 
0& 0& \frac{ \mu_{S}}{2}& 0 & 0 & 0 & 0
\end{array}
\right),
\end{eqnarray}
where we have defined
\begin{eqnarray}
A&=& \left( \frac{f_{1}v}{3}+ \frac{f_{2}z}{3} \right), \nonumber \\
B&=& \left( \frac{f^{\prime}_{1}v^{\prime}}{3}+ \frac{f^{\prime}_{2}z^{\prime}}{3} 
\right).
\end{eqnarray}

Using Eqs.(\ref{vcp},\ref{pdmsusy331rc},\ref{pamsusy331rc}), we get the result presented in 
Figs.(), where we can note, there are 
the following superior limit in (GeV) 
\begin{itemize}
\item For the lighest double chargino $\tilde{\chi}^{0}_{1}$
\begin{equation}
m \leq 65 \,\ {\mbox GeV},
\end{equation}
\item For $\tilde{\chi}^{0}_{2}$
\begin{equation}
m \leq 75.\,\ {\mbox GeV},
\end{equation}
\end{itemize}
these results are in agreement with the experimental limits \cite{GAMBIT:2018gjo}, see also our Eq.(\ref{limitesnoMSSM}).

\section{Conclusions}
\label{sec:conclusion}
We have studied the fermions masses of supersymmetric 
331 model with the sextet and anti-sextet and our results are in 
agreement with the actual experimental data. Here we consider 
only the anti-sextet $S$ get VEV. We want to extend this analysis by 
also allowing the sextet $S^{\prime}$ to acquire vev, in similar way as 
we have done recently \cite{Rodriguez:2022hsj,Rodriguez:2024rvf}. 

\begin{center}
{\bf Acknowledgments} 
\end{center}
We would like thanks V. Pleitez for useful discussions above 331 models. We 
also to thanks IFT for the nice hospitality during my several visit 
to perform my studies about the severals 331 Models and also for done 
this article.


\end{document}